\documentclass[aps,prd,twocolumn,showpacs,nofootinbib,showkeys]{revtex4-1}

\usepackage{amssymb} \usepackage{amsmath} \usepackage{graphicx}
\usepackage{epsfig,latexsym}

\RequirePackage{xspace} \allowdisplaybreaks

\begin{document}

\def\bef{\begin{figure}}
\def\eef{\end{figure}}

\newcommand{\nl}{\nonumber\\}

\newcommand{\ans}{ansatz }
\newcommand{\be}[1]{\begin{equation}\label{#1}}
\newcommand{\beq}{\begin{equation}}
\newcommand{\ee}{\end{equation}}
\newcommand{\beqn}[1]{\begin{eqnarray}\label{#1}}
\newcommand{\eeqn}{\end{eqnarray}}
\newcommand{\bd}{\begin{displaymath}}
\newcommand{\ed}{\end{displaymath}}
\newcommand{\mat}[4]{\left(\begin{array}{cc}{#1}&{#2}\\{#3}&{#4}
\end{array}\right)}
\newcommand{\matr}[9]{\left(\begin{array}{ccc}{#1}&{#2}&{#3}\\
{#4}&{#5}&{#6}\\{#7}&{#8}&{#9}\end{array}\right)}
\newcommand{\matrr}[6]{\left(\begin{array}{cc}{#1}&{#2}\\
{#3}&{#4}\\{#5}&{#6}\end{array}\right)}
\newcommand{\cvb}[3]{#1^{#2}_{#3}}
\def\lsim{\raise0.3ex\hbox{$\;<$\kern-0.75em\raise-1.1ex
e\hbox{$\sim\;$}}}
\def\gsim{\raise0.3ex\hbox{$\;>$\kern-0.75em\raise-1.1ex
\hbox{$\sim\;$}}}
\def\abs#1{\left| #1\right|}
\def\simlt{\mathrel{\lower2.5pt\vbox{\lineskip=0pt\baselineskip=0pt
           \hbox{$<$}\hbox{$\sim$}}}}
\def\simgt{\mathrel{\lower2.5pt\vbox{\lineskip=0pt\baselineskip=0pt
           \hbox{$>$}\hbox{$\sim$}}}}
\def\unity{{\hbox{1\kern-.8mm l}}}
\newcommand{\eps}{\varepsilon}
\def\ep{\epsilon}
\def\ga{\gamma}
\def\Ga{\Gamma}
\def\om{\omega}
\def\omp{{\omega^\prime}}
\def\Om{\Omega}
\def\la{\lambda}
\def\La{\Lambda}
\def\al{\alpha}
\newcommand{\ov}{\overline}
\renewcommand{\to}{\rightarrow}
\renewcommand{\vec}[1]{\mathbf{#1}}
\newcommand{\vect}[1]{\mbox{\boldmath$#1$}}
\def\tm{{\widetilde{m}}}
\def\mcirc{{\stackrel{o}{m}}}
\newcommand{\Dm}{\Delta m}
\newcommand{\dm}{\varepsilon}
\newcommand{\tanb}{\tan\beta}
\newcommand{\nbar}{\tilde{n}}
\newcommand\PM[1]{\begin{pmatrix}#1\end{pmatrix}}
\newcommand{\up}{\uparrow}
\newcommand{\down}{\downarrow}
\def\omE{\omega_{\rm Ter}}
%
%%%%%%%%%%     mauri    %%%%%%%%%%%%%%%%%%%%%%%%%%%%%%%%%

\newcommand{\Dsusy}{{susy \hspace{-9.4pt} \slash}\;}
\newcommand{\DCP}{{CP \hspace{-7.4pt} \slash}\;}
\newcommand{\mc}{\mathcal}
\newcommand{\gr}{\mathbf}
\renewcommand{\to}{\rightarrow}
\newcommand{\gtc}{\mathfrak}
\newcommand{\wh}{\widehat}
\newcommand{\br}{\langle}
\newcommand{\kt}{\rangle}

%%%%%%%%%%%%%%%%%%%%%%%%%%%%%%%%%%%%%%%%%%%%%%%%%%%%%%%%%%

% barbara Ricci  %definizione di minore e maggiore simile
\def\lsim{\mathrel{\mathop  {\hbox{\lower0.5ex\hbox{$\sim$}
\kern-0.8em\lower-0.7ex\hbox{$<$}}}}}
\def\gsim{\mathrel{\mathop  {\hbox{\lower0.5ex\hbox{$\sim$}
\kern-0.8em\lower-0.7ex\hbox{$>$}}}}}
%%%%%%%%%%%%%%%%%%%%%%%%%%%%%%%%%%

\def\nn{\\  \nonumber}
\def\de{\partial}
\def\brf{{\mathbf f}}
\def\bbf{\bar{\bf f}}
\def\bF{{\bf F}}
\def\bbF{\bar{\bf F}}
\def\bA{{\mathbf A}}
\def\bB{{\mathbf B}}
\def\bG{{\mathbf G}}
\def\bI{{\mathbf I}}
\def\bM{{\mathbf M}}
\def\bY{{\mathbf Y}}
\def\bX{{\mathbf X}}
\def\bS{{\mathbf S}}
\def\bb{{\mathbf b}}
\def\bh{{\mathbf h}}
\def\bg{{\mathbf g}}
\def\bla{{\mathbf \la}}
\def\bmu{\mathbf m }
\def\by{{\mathbf y}}
\def\bmu{\mbox{\boldmath $\mu$} }
\def\bsig{\mbox{\boldmath $\sigma$} }
\def\bunity{{\mathbf 1}}
\def\cA{{\cal A}}
\def\cB{{\cal B}}
\def\cC{{\cal C}}
\def\cD{{\cal D}}
\def\cF{{\cal F}}
\def\cG{{\cal G}}
\def\cH{{\cal H}}
\def\cI{{\cal I}}
\def\cL{{\cal L}}
\def\cN{{\cal N}}
\def\cM{{\cal M}}
\def\cO{{\cal O}}
\def\cR{{\cal R}}
\def\cS{{\cal S}}
\def\cT{{\cal T}}
\def\eV{{\rm eV}}
%
%%%%%%%%%%%%%%%%%%%%%%%%%%%%%%%%%%%%%

\title{Testing B-violating signatures from Exotic Instantons in future colliders}

\author{Andrea Addazi$^1$}\email{andrea.addazi@infn.lngs.it}
\author{Xian-Wei Kang$^{2,3}$}\email{kxw198710@126.com}
\author{Maxim Yu. Khlopov$^4$}\email{khlopov@apc.in2p3.fr}
\affiliation{$^1$Dipartimento di Fisica,
 Universit\`a di L'Aquila, 67010 Coppito AQ and
LNGS, Laboratori Nazionali del Gran Sasso, 67010 Assergi AQ, Italy\\
$^2$Institute of Physics, Academia Sinica, Taipei, Taiwan 115\\
$^3$Institute for Advanced Simulation, J\"ulich Center for Hadron
Physics, and Institut f\"ur
Kernphysik, Forschungszentrum J\"ulich, D-52425 J\"ulich, Germany\\
$^4$Centre for Cosmoparticle Physics Cosmion; National Research
Nuclear University MEPHI (Moscow Engineering Physics Institute),
Kashirskoe Sh., 31, Moscow 115409, Russia; APC laboratory 10, rue
Alice Domon et L\'eonie Duquet 75205 Paris Cedex 13, France}

\begin{abstract}

We discuss possible implications of Exotic Stringy instantons
for Baryon-violating signatures in future colliders.
In particular, we discuss
 high-energy quarks collisions and $\Lambda-\bar{\Lambda}$ transitions.
In principle, $\Lambda-\bar{\Lambda}$ process
can be probed by high-luminosity electron-positron colliders.
However, we find
that an extremely high luminosity is needed in order to provide a
(somewhat) stringent bound compared to the current data on
 $NN\rightarrow \pi\pi,KK$.
On the other hand, (exotic) instanton-induced six quark interactions can be tested
in near future high-energy colliders
beyond LHC, at energies around $20-100\, \rm TeV$.
Super proton-proton collider (SppC) is capable of such measurement
given the proposed energy level of 50-90 TeV.
Comparison with other channels is made.
In particular, we show the compatibility of our model
with neutron-antineutron and $NN\rightarrow \pi\pi,KK$ bounds.

\end{abstract}



\maketitle

\section{Introduction}

In recent companion papers,
implications of exotic stringy instantons in
$B-L$ violating rare processes and baryogenesis
were explored
\cite{Addazi:2014ila,Addazi:2015ata,Addazi:2015rwa,Addazi:2015hka,Addazi:2015eca,Addazi:2015fua,Addazi:2015oba,Addazi:2015goa,Addazi:2015ewa,Addazi:2015yna,Addazi:2016xuh,Addazi:2016mtn}

As known, in open string theories, instantons are
 (Euclidean) D-branes wrapping n-cycles on the Calabi-Yau (CY) manifold.
  These solutions were calculated and classified
in the literature (see \cite{Bianchi:2009ij} for a complete review on these subjects).

In this paper, we will suggest that the exotic instantons can be tested
in collider physics. As shown in \cite{Addazi:2015goa} six quarks $\Delta B=2$
violating transitions can be generated by one exotic instanton
in the context of a low scale string theory scenario with $M_{S}\simeq 10\div 100\, \rm TeV$.
The model suggested in \cite{Addazi:2015goa} is theoretically motivated by
baryogenesis and the hierarchy problem of the Higgs mass, while
embedded a theory of quantum gravity.
As regards to baryogenesis, the tunneling probability of $B-L$ violating processes
can be enhanced in the thermal bath, as it happens for standard model electroweak sphalerons
 \cite{Kuzmin:1985mm,Kuzmin:1987wn}.
This implies a $B-L$ first order phase transition which could explain the observed matter-antimatter
asymmetry.
On the other hand, a low string scale theory can alleviate the strong hierarchy problem of the Higgs mass
of $17$ orders, reducing the hierarchy to a small number (comparable to the Yukawa coupling of the electron).
The possibility to test exotic instantons in collider may allow
to the test low scale string theory in the fully non-perturbative regime.
This may be insightfully important also to understand the
issues from geometric moduli stabilization in string compactification.
In fact, string instantons and string fluxes may generate non-perturbative effective potentials
stabilizing the geometric moduli  and allowing for a {\it string vacuum safety}.

In particular, we will show how exotic instantons
can generate $\Lambda-\bar{\Lambda}$ transitions,
and $qq \rightarrow \bar{q}\bar{q}\bar{q}\bar{q}$
$qq\rightarrow \bar{\tilde{q}}\bar{\tilde{q}}\bar{\tilde{q}}\bar{\tilde{q}}$
 high-energy collisions.
Electron-positron colliders with luminosity much higher than Belle and BaBar
could be used to (indirectly) test such a scenario.
Indeed, we find that an extremely high luminosity is needed,
 and even that look unrealizable in the near future.
On the other hand, we argue how
compared measured in neutron-antineutron physics
and in high energy colliders beyond LHC
can provide tests for our model
in the near future. In this sense, the high energy frontier is the preferred
experimental direction of our model, with respect to the high luminosity one.

The letter is organized as following:
In Sec.~\ref{sec2} we discuss our theoretical model,
in Sec.~\ref{sec3} its phenomenology in electron-positron colliders,
and in Sec.~\ref{sec4} its phenomenology and parameter space
in comparison with several other different possible channels before
our conclusions.

\section{Theoretical side}\label{sec2}

 $B,L$ number conservations can be {\it dynamically} violated by
 non-perturbative quantum gravity effects known as
 {\it exotic stringy instantons}.
 Exotic stringy instantons are
 Euclidean D-branes (or E-branes),
 intersecting physical D-brane stacks.
In our letter, we will consider IIA string-theories \footnote{For classic papers on open string theories see
\cite{Sagnotti0,Sagnotti1,Pradisi,Sagnotti4,Sagnotti5}. In open string theories, calculations of scattering
amplitudes are much simpler than F-theories or heterotic string theories.
}. In this case, "exotic instantons"
 are $E2$-branes, wrapping different
 3-cycles on a Calabi-Yau compactification with respect to physical $D6$-branes.
 A MSSM can be embedded in a quiver theory
 with three or more nodes.
 In the low energy limit, these quivers can produce gauge theories
 $U(3)_{a}\times U(2)_{b}\times U(1)_{c}$
  or
  $U(3)_{a}\times Sp(2)_{b}\times U(1)_{c}\times U'(1)_{s}$ or
  $U(3)_{a}\times U(2)_{b}\times U(1)_{c}\times U'(1)_{d}$ or eventually higher node extensions
  \footnote{See  \cite{E1,E4,E7,DB2,DB10,DB19,DB25}
  for (incomplete) litterature on intersecting D-branes models. }.
   Let us remark that the basic fundamental elements
are: i) $D6$-branes wrapping 3-cycles in the Calabi-Yau
$CY_{3}$; ii) $\Omega$-planes; iii) $E2$-instantons;
iv)
open strings \footnote{Before the orientifold projection,
in order to restore the correct balance of arrows
one has also to introduce flavor branes for some of these models.
See
\cite{Bianchi:2013gka} for a discussion of Unoriented quivers with Flavour branes. }.

Open strings are attached to D6 and E2-branes.
Let us also recall that open (un)oriented strings
attached to two intersecting D6-brane stacks
will reproduce (MS)SM matter fields as lower energy excitations,
in the limit  $\alpha_{s}'\rightarrow 0$.
The number of intersections among the stacks
will correspond to the number of generations.
As an example, the three generations of lepton superfields
$L$ comes from open strings attached
to $U(2)$ or $Sp(2)$ stack and
a $U(1)$ stack, with these stacks intersecting three times each other.
The hypercharge $U(1)_{Y}$ is
reconstructed as a massless linear combinations of
the $U(1)$'s in the model.
For example for $U(3)_{a}\times U(2)_{b}\times U(1)_{c}$,
\be{hypercharge}
Y=q_{a}Q_{a}+q_{b}Q_{b}+q_{c}Q_{c}\,,
\ee
where $Q_{a,b,c}$ corresponds to
$U(1)_{a}\subset U(3)_{a}$ and $U(1)_{b} \subset U(2)_{b}$
and $U(1)_{c}$ respectively.
On the other hand, two linear independent combinations
of $U(1)$'s  orthogonal to (\ref{hypercharge})
will be anomalous \footnote{In string theory,
anomalous $U(1)$s can be cured
through the
generalized Green-Schwarz mechanism.
The two anomalous vector bosons $Z',Z''$
get a mass of the order of the String scale,
through St\"uckelberg mechanism.
Peculiarly, they have to interact
with $Z,\gamma$
through generalized Chern-
Simon (GCS) terms in order to have
 anomalies' cancellations.
See \cite{Stuck1,Stuck11} for an extensive
discussion on these aspects. }.
After this short introduction,
let us consider, the presence of an $E2$-brane
intersecting two times the $U(3)$-stack,
two times the $U(1)$-stack, four time the $\hat{U}(1)$-stack (image of $U(1)$ with respect to $\Omega$).
For $\alpha_{s}'\rightarrow 0$, this construction generate ,
effective Lagrangian terms among
ordinary superfields $U^{c},D^{c}$ and
fermionic moduli (also called modulini):
\be{effectivemod}
\mathcal{L}_{eff}\sim k_{f}^{(1)}U_{f}^{i}\tau_{i}\alpha+k_{f}^{(2)}D_{f}^{i}\tau_{i}\beta\,,
\ee
where $\tau^{i}$ modulini live at $U(3)$-$E2$ intersections,
$\alpha$ modulini at $U(1)-E2$ and $\beta$ at $\hat{U}(1)-E2$.
Here, we consider
an $E2$-instanton with
a Chan-Paton factor $O(1)$.
As prescribed by  instanton calculation,
we will
integrate out modulini at the D6-E2 intersections,
and we will obtain
\beqn{obtain}
\mathcal{W}_{E2-D6-\hat{D6}}&=&\int d^{6}\tau d^{4}\beta d^{2}\alpha e^{-\mathcal{L}_{eff}}\nl
&=&\mathcal{Y}^{(1)}\frac{e^{-S_{E2}}}{M_{S}^{3}}\epsilon_{ijk}\epsilon_{i'j'k'}U^{i}D^{j}D^{k}U^{i'}D^{j'}D^{k'}\nl,
\eeqn
where $\mathcal{Y}^{(1)}_{f_{1}f_{2}f_{3}f_{4}f_{5}f_{6}}=k^{(1)}_{f_{1}}k^{(1)}_{f_{2}}k^{(2)}_{f_{3}}k^{(2)}_{f_{4}}k^{(2)}_{f_{5}}k^{(2)}_{f_{6}}$ is the flavor matrix determined by the couplings $k^{(1,2)}$ derived from mixed disk amplitudes.
 We can assume these as free parameters, parametrizing our ignorance about
 the particular geometry of the E2-instanton considered.
 A Superpotential (\ref{obtain})
 can generate diagrams like the one in Fig.2.
 In particular, $n-\bar{n}$ and $\Lambda-\bar{\Lambda}$ transitions
 are generated by the two effective operators
\beqn{ope}
\mathcal{O}_{n\bar{n}}+\mathcal{O}_{\Lambda\bar{\Lambda}}&=&\frac{y_{1}}{\mathcal{M}_{E2}^{3}M_{SUSY}^{2}}(u^{c}d^{c}d^{d})(u^{c}d^{c}d^{c})\nl
&& +\frac{y_{2}}{\mathcal{M}_{E2}^{3}M_{SUSY}^{2}}(u^{c}d^{c}s^{c})(u^{c}d^{c}s^{c})\,,
\eeqn
where $\mathcal{M}_{E2}^{3}=e^{+S_{E2}}M_{S}^{3}$,
$y_{1}=\mathcal{Y}^{(1)}_{111111}$ and $y_{2}=\mathcal{Y}^{(1)}_{112112}$.
More details and analysis of
explicit quivers were extensively discussed in
\cite{Addazi:2015goa}. These operators
correspond to the generation of an effective Majorana
mass for the neutron and for the $\Lambda$ baryon.

$e^{-S_{E2}}$ is the effective action of $E2$-brane
wrapping 3-cycles on the $CY_{3}$.
It is
related to the string coupling as
\beqn{rel}
e^{-S_{E2}}=e^{-\mathcal{V}_{E2}/g_{s}+   i \sum_{r}q_{r}a_{r}}\,,
\eeqn
where $\mathcal{V}_{E2}$ is the volume wrapped by the $E2$-brane
and the imaginary part consists of a sum all over Ramond-Ramond axions.
For $l>l_{S}$, Exotic instantons have to respect the universal string theory bound on non-perturbative effects:
\beqn{bound}
|e^{-S_{E2}}|\leq e^{-\frac{2}{g_{s}}}
\eeqn
interpreted as a bound from instanton-antinstanton virtual pair diagram,
{\it i.e} as a bound on their partition function.
This bound is very important for the considerations in the following.
In fact, $g_{s}<<1$ will suppress $E2$-transitions.
On the other hand, scenari in which $g_{s}=0.25 \div 1$
suggest a coupling strong as $e^{-10}\div e^{-2}\simeq 5\times 10^{-5}\div 1.1\times 10^{-1}$,
with a small volume $\mathcal{V}_{E2}$
This result is peculiarly different with respect to non-perturbative classical configurations
in field theories, like sphalerons, usually suppressed as $e^{-1/g_{YM}^{2}}$.
In particular, for electroweak gauge instantons, the suppression factor can be proportional to $e^{-10^{4}}$ or so.
A so high string scale as the one desired in our case
has to be consistently compatible with Yang-Mills coupling
and $M_{Pl}/M_{S}$ ratio.
Let us remind that, by dimensional reduction to 4d,
\be{gYM}
\frac{1}{g_{YM}^{2}}=\frac{M_{S}^{3}\mathcal{V}_{3}}{(2\pi)^{4}g_{s}}\,,
\ee
\be{Mpl}
M_{Pl}^{2}=\frac{M_{S}^{8}\mathcal{V}_{6}}{(2\pi)^{7}g_{s}^{2}}\,,
\ee
where $\mathcal{V}_{3}$ is the volume wrapped by $D6$-branes (stacks)
corresponding to YM bosons; $\mathcal{V}_{6}$ is the total internal volume.
The hierarchy among YM couplings and a high string coupling
is understood as a volume suppression hierarchy.
Let us note that generically the volume of three-cycles
are not directly related to the total internal volume,
even if in simple compactifications like isotropic toroidal ones they are related
as $\mathcal{V}_{3}\sim \sqrt{\mathcal{V}_{6}}$.
This allows more variability among String scale and Planck scale
hierarchies.

On the other hand, the suppression factor $e^{-S_{E2}}$ can also be compensated
by coefficients in $k^{(1)}_{f}$ of mixed disk amplitudes.
These coefficients can be higher than one and their combinations
can give rise to an appreciable enhancement of $\mathcal{Y}^{(1)}$ flavor components
\footnote{However, let us mention that recent trends in non-perturbative string theory and string phenomenology
suggest that for distances of $l<l_{S}$ 3-cycle volume factor can be
collapsed on the $CY_{3}$ singularity \cite{collapse1,collapse2}.
In this regime, saddle point approximation cannot be trusted anymore
so that our previous estimates are not more valid in this case.
In this case an enhancement of exotic instanton processes is expected.}.

\subsection{Exotic instantons in high energy collisions}

In this section, we will consider two quarks
high energy collisions induced by exotic instantons.
For $s<<\Lambda^{2}$, the scattering amplitude is
just reduced to a contact six quark interaction
\beqn{pointlike}
\mathcal{A}(q_{f_{1}}^{c}q_{f_{2}}^{c}\rightarrow \bar{q}_{f_{3}}^{c}\bar{q}_{f_{4}}^{c}\bar{q}_{f_{5}}^{c}\bar{q}_{f_{6}}^{c})\simeq \mathcal{Y}^{(1)}_{f_{1}f_{2}f_{3}f_{4}f_{5}f_{6}}\frac{\Lambda^{-3}}{M_{SUSY}^{2}}
\eeqn
with $q^{c}=u^{c},d^{c},....t^{c}$ RH quarks and $\Lambda^{-3}\simeq M_{S}^{-3}e^{-S_{E2}}$.
A less suppressed channel is $q^{c}q^{c}\rightarrow \bar{\tilde{q}}^{c}\bar{\tilde{q}}^{c}\bar{\tilde{q}}^{c}\bar{\tilde{q}}^{c}$.
In the low energy limit, its amplitude is
\beqn{pointlike}
\mathcal{A}(q_{f_{1}}^{c}q_{f_{2}}^{c}\rightarrow \bar{\tilde{q}}_{f_{3}}^{c}\bar{\tilde{q}}_{f_{4}}^{c}\bar{\tilde{q}}_{f_{5}}^{c}\bar{\tilde{q}}_{f_{6}}^{c})\simeq \mathcal{Y}^{(1)}_{f_{1}f_{2}f_{3}f_{4}f_{5}f_{6}}\Lambda^{-3}
\eeqn
The corresponding quarks-squarks cross section
can be evaluated by integrating the squared of the amplitude all over the 4-dimensional
phase space $d\Phi_{4}$.
We define the Mandelstam variables
$s_{12}=(p_{1}+p_{2})^{2}$,
$s_{34}=(p_{3}+p_{4})^{2}$,
where $p_{1,2,3,4}$ are squark momenta (final states)
and $s=E_{CM}^{2}$.
One finds that
\beqn{correct}
d\sigma_{q_{f_{1}}^{c}q_{f_{2}}^{c}\rightarrow \bar{\tilde{q}}_{f_{3}}^{c}\bar{\tilde{q}}_{f_{4}}^{c}\bar{\tilde{q}}_{f_{5}}^{c}\bar{\tilde{q}}_{f_{6}}^{c}}=d\hat{\Omega} \frac{\mathcal{C}}{4\Lambda^{6}}\mathcal{P}_{2}(s,s_{12},s_{34})\,,
\eeqn
where
\beqn{factor1}
\mathcal{C}=\frac{C_{4}}{(8\pi^{2})^{3}} Tr\left\{\mathcal{Y}^{(1)^{\dagger}}\mathcal{Y}^{(1)} \right\}\,,
\eeqn
with $C_{4}$ a combinatorial factor depending on the number of equal particles in the final state,
and $d\hat{\Omega}$ the integral all over scattering angles
\beqn{angular}
d\hat{\Omega}=d\cos \theta\frac{d\phi}{2\pi} d\cos \hat{\theta}_{12}\frac{d\hat{\phi}_{12}}{2\pi}d\cos \hat{\theta}_{34}\frac{d\hat{\phi}_{34}}{2\pi}\,.
\eeqn
The notation $\hat{\phi}_{ij}$ indicates variables evaluated with respect to the rest frame of $q_{ij}$.
$\mathcal{P}_{2}$ is a complicated polynomia of $s,s_{12},s_{34}$ and subleading logaritimic functions of
expression later:
$$\mathcal{P}_{2}(x,y,z)=-\zeta_{1}(y)\zeta_{1}(z)[-4 y \log\zeta_{2}^{+}(x,y,z) $$
$$+x\log\zeta_{2}^{-}(x,y,z)+y^{2}+\zeta_{3}(x,y,z)]$$
where
$$\zeta_{1}(x)=\sqrt{1-\frac{2(m_{1}+m_{2})}{x}+\frac{(m_{1}-m_{2})^{2}}{x^{2}}}$$
$$\zeta_{2}^{\pm}(x,y,z)=-\frac{x^{2}-2x(x+y)+(y-z)^{2}-x\pm y \mp z}{\Lambda^{2}}$$
$$\zeta_{3}(x,y,z)=\frac{1}{3}\left[x^{2}+5x(y+z)- \frac{2(y-z)^{2}}{x}\sqrt{\zeta_{2}^{+}+x-y+z}\right]$$
with $m_{1,2,3,4}$ squark masses in the final state.

For $s\rightarrow  \Lambda^{2}$,  cross sections rapidly grows up.
$\bar{s}=\Lambda^{2}$ corresponds to the bound of the unitarity break-down.
In other word, our effective cross-section would badly breaks the Froissart bound.

For $s\simeq \Lambda^{2}$, the contact interaction approximation looses validity:
scatterings are probing the fully non-perturbative regime:
an exotic instanton is a fully non-perturbative configuration.
Resonant production at the $s\simeq \Lambda^{2}$
corresponds to an infinite series of stringy amplitudes with six open-strings' insertions
$\sum_{g=0}^{\infty}\mathcal{A}^{g}(z_{1},z_{2},z_{3},z_{4},z_{5},z_{6})$,
where g is the genus (loops).
This is a technical problem   in the high energy limit common  to all non-perturbative (euclidean) configurations,
like QCD instantons, sphalerons and so on.
However, let us note that
in our model $\Lambda =e^{+S_{E2}} M_{S}>M_{S}$. As a consequence, an infinite tower of stringy higher spins states
are excited at $\Lambda$ and they will unitarize the S-matrix.
This is a generic conclusion coming from unitarization arguments of
string theory amplitudes, also connecting to the CPT symmetry in string theory.
We consider the problem in the fixed angle kinematical regime,
which is also particularly suitable with the experimental set-up of colliders.
We argue that fixed angle scattering amplitudes of (six) open strings are expected to
exponentially fall down with energy as
\beqn{exp}
\mathcal{A}_{s>M_{S}^{2}}^{tree-level}=\mathcal{A}^{QFT}e^{-\alpha' s \log \alpha' s+...}
\eeqn
while N-loops amplitudes behaves as \footnote{The result found by Gross-Mende in \cite{Gross:1987ar}
is valid for four open string amplitude. Their result is
expected to be qualitatively the same as for six points amplitudes
in $2\rightarrow 4$ fixed angle scatterings.   } \cite{Gross:1987ar}
\beqn{Nexp}
\mathcal{A}^{N-loops}_{s>M_{S}^{2}}=\mathcal{A}^{QFT}e^{-\frac{1}{N}\alpha' s \log \alpha' s+...}\,.
\eeqn
In Fig.1, we show the qualitative universal part  (flavor and combinatory independent) of two quark four squark cross section,
assuming squark masses smaller than $\Lambda$-scale.
For $E_{CM}\simeq \Lambda$ the process lies into the
 fully non-perturbative stringy regime.
From the physical point of view, we expect that, with the growing of CM energy, the $E2$-brane starts to oscillate
and its dynamics is described by oscillations of modulini fields.
The number of modulini is a topological invariant of the exotic instanton solution, associated with the invariance of intersections
with physical D-branes. At the non-perturbative scale, open strings associated to modulini
have to reggeize. Loop-corrections to tree-level mixed-disk amplitudes
at fixed angle are expected to add exponentially suppressed contributions
in the form of Eq.(\ref{Nexp}).
We conjecture that this is the main contribution to
the unitarization of the scattering amplitude for fixed angle kinematical regime
and $s>>\Lambda^{2}$.

\begin{figure}[t]
\centerline{ \includegraphics [height=5cm,width=0.8\columnwidth]{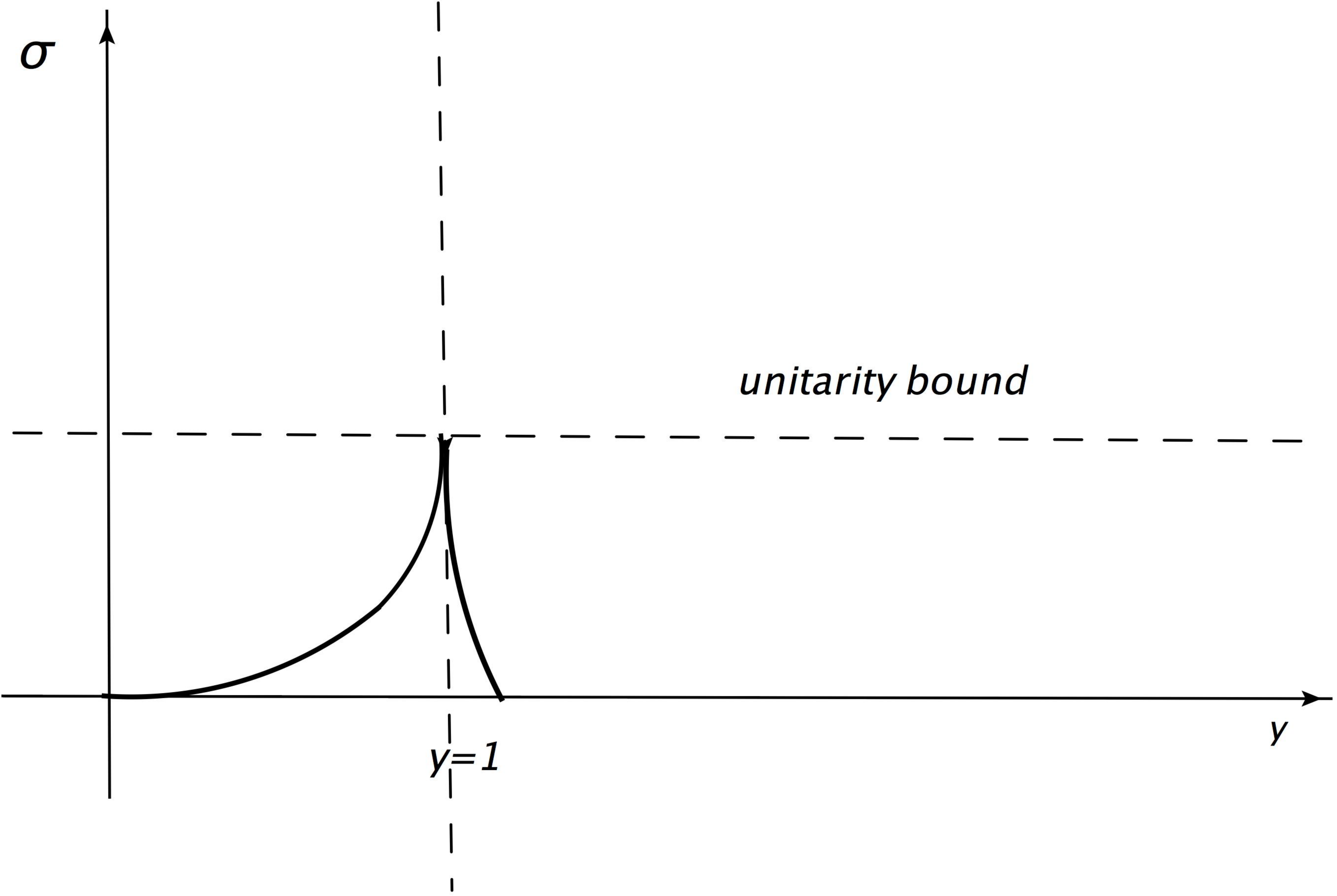}}
\vspace*{-1ex}
\caption{We show the two quark four squarks cross section $\sigma(y)$ (flavor independent part) with respect to the normalized s-variable $y=s/\Lambda^{2}$
and neglecting squark masses.
For $y<1$, $\sigma\sim y^{2}$; for $y\simeq 1$ the amplitude will saturate the unitarity bound
of string theory. For $y>1$ and fixing the scattering angle,  the amplitude falls off exponentially with the production of the non-perturbative configuration --as displayed in this figure. }
\label{plot}   % \ref{plot}
\end{figure}

We will return to phenomenological implications in the next sections.

\section{Phenomenology of $\Lambda-\bar{\Lambda}$ transitions}\label{sec3}

\begin{figure}[t]
\centerline{ \includegraphics [height=4.5cm,width=0.5\columnwidth]{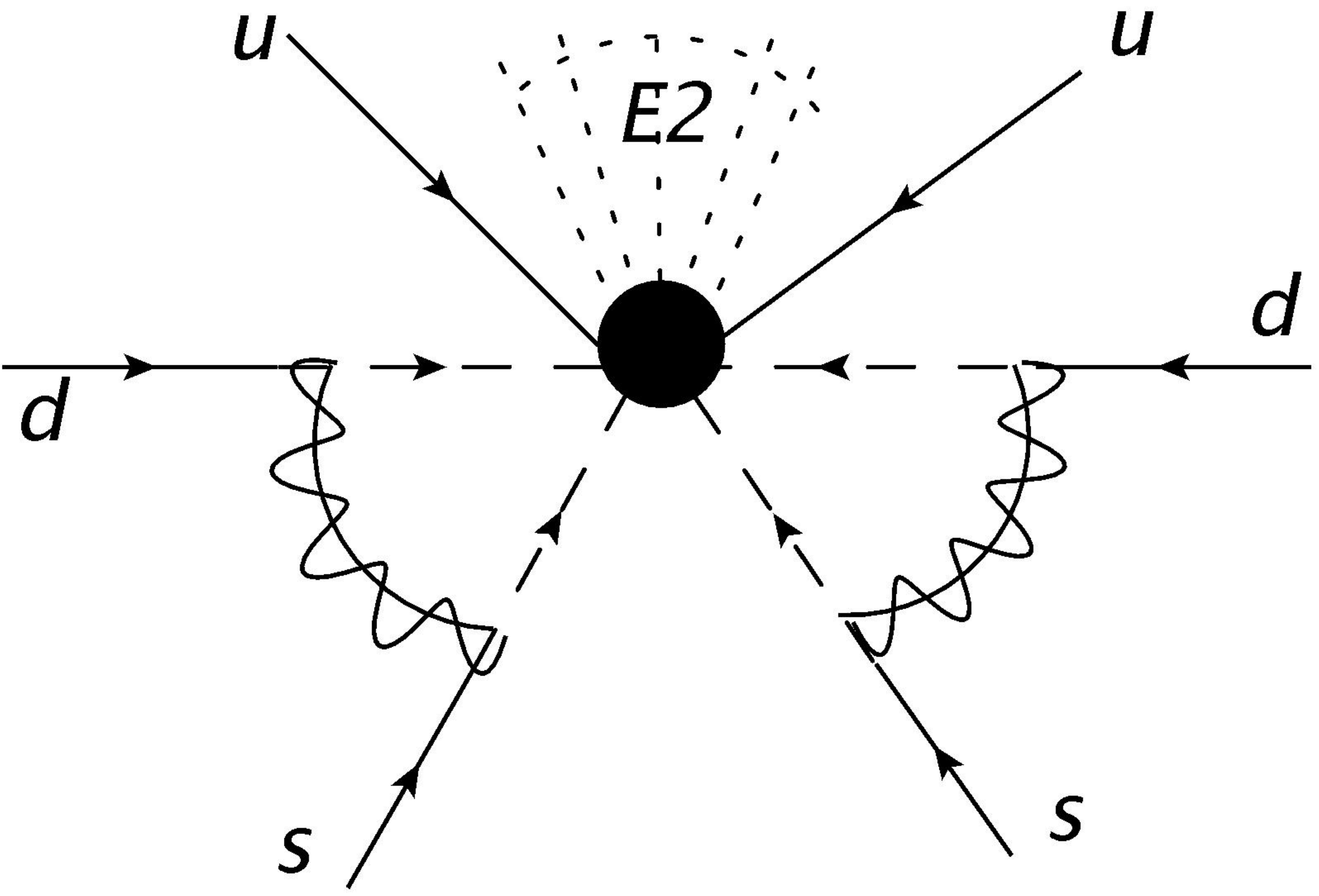}}
\vspace*{-1ex}
\caption{Diagram for a $\Lambda-\bar{\Lambda}$ transition.
$u,d,s$ are right-handed up, down and strange quarks.
Four quarks in the diagram are converted to the
corresponding squarks through the
exchange of two gaugini.
The black vertex is
directly generated by an Exotic Instanton.}
\label{plot}   % \ref{plot}
\end{figure}

Now let us discuss the implications of $\Lambda-\overline\Lambda$
oscillations on the phenomenological side \footnote{Of course, in the standard model analysis, these baryon-violating
effects need not to be considered, e.g., see the $CP$ violation of $\Lambda_c\to\Lambda \pi$ decay \cite{Lambdac} and also the conventional
$N\bar N$ scattering \cite{KangHM1,KangHM2,KangHM3,KangHM4}.}.
So far that no baryon number violating process
has been observed in the Standard Model (SM), thus any possible signal would be definitely exciting and efforts along this direction are deserved.
New physics effects can be also probed at hadron-hadron colliders. As discussed in Ref.~\cite{Kang}, BES detector with huge $J/\psi$ data sample has been proposed
to measure the possible $\Lambda-\overline\Lambda$ oscillation by studying the quantum coherent $\Lambda\overline\Lambda$ states. Following the treatment of
$D^0-\overline D^0$ mixing \cite{PDGrev}. To avoid the complications due to mixing, the charged $B$ was used as an example in Ref.~\cite{Bl4}. The eigenstates emerging from
the oscillations can be written as
\begin{eqnarray}
|\Lambda_H\rangle=p|\Lambda\rangle+q|\overline\Lambda\rangle,\nonumber\\
| \Lambda_L\rangle=p|\Lambda\rangle-q|\overline\Lambda\rangle,
\end{eqnarray}
with the normalization condition $|p^2|+|q|^2=1$. The subscript ``H'' (``L'') means the heavy (light) mass eigenstates. Here $p$ and $q$
can be parametrized as
 $\sqrt{1+z}/\sqrt{2}$ and $\sqrt{1-z}/\sqrt{2}$, respectively, where $z$ involves the oscillation mass $\delta m_{\Lambda\overline\Lambda}$ appearing
in the off-diagonal elements of the effective Hamiltonian used to describe the time evolution of the $\Lambda-\overline\Lambda$ system. For details, see Ref.~\cite{Kang} or the general consideration in \cite{PDGrev}. One can define the following mass and width:
\begin{eqnarray}
m\equiv \frac{m_H+m_L}{2},\quad \Delta m\equiv m_H-m_L\,\nonumber\\
\Gamma\equiv \frac{\Gamma_H+\Gamma_L}{2},\quad \Delta\Gamma\equiv\Gamma_H-\Gamma_L,
\end{eqnarray}
from which the mixing parameters are defined as
\begin{equation}
x_\Lambda\equiv\frac{\Delta m}{\Gamma},\quad y_\Lambda\equiv \frac{\Delta\Gamma}{2\Gamma}.
\end{equation}
For the free $\Lambda$ case without external magnetic fields, $\Delta m=2\delta m_{\Lambda\overline\Lambda}$. For simplicity, we will confine ourselves
to this case. The influence of external magnetic field is discussed in Ref.~\cite{Kang}.
The dominant decay mode of $\Lambda$ is $\Lambda\to p\pi^-$ with branching ratio $64\%$ \cite{PDG}, and correspondingly
$\overline\Lambda\to\overline p\pi^+$ assuming $CP$ invariance.
In the case that $\Lambda-\overline\Lambda$ oscillation happens, the process $\overline\Lambda\to\Lambda\to p\pi^-$ will be possible.
This phenomenon can be probed by counting the event number $\mathcal N$ for $J/\psi\to\Lambda\overline\Lambda\to(p\pi^-)(p\pi^-)$ --wrong-sign decay-- while
the right-sign decay would be $J/\psi\to\Lambda\overline\Lambda\to(p\pi^-)(\overline p\pi^+)$. Time-integrated decay rate for the wrong-sign decay relative
to the right sign is found to be
\begin{equation}\label{ratio}
\mathcal R=\frac{\mathcal N \Big(J/\psi\to\Lambda\overline\Lambda\to(p\pi^-)(p\pi^-)\Big)}{\mathcal N \Big(J/\psi\to\Lambda\overline\Lambda\to(p\pi^-)(\overline p\pi^+)\Big)}
=\frac{x_\Lambda^2+y_\Lambda^2}{2}.
\end{equation}
A discussion of time-dependent observables is also presented in \cite{Kang}, and the BES-III detector is capable to access this time information.
Assuming $y_\Lambda=0$ -- it is indeed a quite small quantity -- one can get an estimate of the mixing mass parameter $\delta m_{\Lambda\overline\Lambda}$ from
Eq.~\eqref{ratio} as
\begin{equation}\label{osc-lum}
\delta m_{\Lambda\overline\Lambda}=\frac{1}{\sqrt{2}}\sqrt{\mathcal R}\Gamma.
\end{equation}
According to the designed luminosity of BEPC-II in Beijing \cite{BESdesign}, $10\times 10^9$ $J/\psi$ and $3\times 10^9$ $\psi'$data samples will be collected by the runnings
per year. If finally no signal of wrong-sign decay can be detected, one should put an upper limit $\mathcal{R}\le 4\times 10^{-7}$ and
correspondingly $\delta m_{\Lambda\overline\Lambda}\le 10^{-15}\,\text{MeV}$ at the 90\% of confidence level (C.L.)  \cite{Kang}, inferring from the
knowledge of interval estimate for very rare signal. Currently, a sample of $1.31\times10^9$ $J/\psi$
events has been collected \cite{Jpsinumber}, then the aforementioned upper limit will be increased by a factor of $\sqrt{10/1.31}\approx2.8$, i.e.
$\delta m_{\Lambda\overline\Lambda}\le 3\times 10^{-15}\,\text{MeV}$. Consequently, the oscillation time will be bounded by the upper limit as $2.5\times10^{-7}$\,s
at 90\% confidence level. This would be the first search in the experiment. See also \cite{Kang}
for more details on this analysis.

Note that $\Upsilon(4S)$ also has the same quantum numbers as $J/\psi$, i.e., $I^G (J^{PC})=0^- (1^{--})$, and can also decay to coherent $\Lambda\overline\Lambda$
states. Belle and BaBar detectors could be used to probe this process as well. Currently, there are $772\times10^6$ $\Upsilon(4S)$ data available in Belle \cite{Belle} and
$471\times10^6$ for BaBar\cite{BaBar}. Taking into account the fact that the non-$B\overline B$ decay mode only constitutes  less than 4\% (at 95\% confidence level)
of the total decay rate, much less $\Lambda\overline\Lambda$ events are expected. Otherwise assuming that most of $\Upsilon(4S)$ decays into $\Lambda\overline\Lambda$, the event number
can increase by one or two orders lager more than BES. Although the above upper limit for $\Lambda-\bar\Lambda$ oscillation could be accessed from the beginning in the experiment,
we require an extremely higher luminosity to get a comparable bound to the $n-\bar n$ case. One can find this point from Eq.~\eqref{osc-lum} -- the oscillation mass is proportional
to the square root of  the luminosity. $n-\bar n$ oscillation time is constrained to be $10^7 \div 10^8$ smaller than the $\Lambda-\bar\Lambda$ oscillation time, and thus the luminosity should be
increased by order $10^{15}$, which seems unrealizable in the near future. In other words, extremely huge data samples would be needed to pin down a more stringent bound on
such new-physics (NP) signal. However, we will show that the required energy to generate such NP phenomenon (e.g., in exotic instanton model below) can  be accessed
in the near future. Recently, the Circular Electron Positron Collider (CEPC) has been proposed in China, which arouses great interests in the
community \cite{CEPC}.
According to the design agenda,
the electron-positron collider will be converted into
a proton-proton collider,
with an unprecedented center-of-mass energy
of 50-90 TeV at the second phase. This is the project of super proton-proton collider (SppC) \cite{Wangyf}. At such a high energy, the new physics beyond
LHC discussed in the present paper would be accessible.
Future measurements are expected to give valuable hints on this research line.

\section{Space of parameters, $n-\bar{n}$ and proton-proton colliders}\label{sec4}

\begin{figure}[t]
\centerline{ \includegraphics [height=8cm,width=0.7\columnwidth]{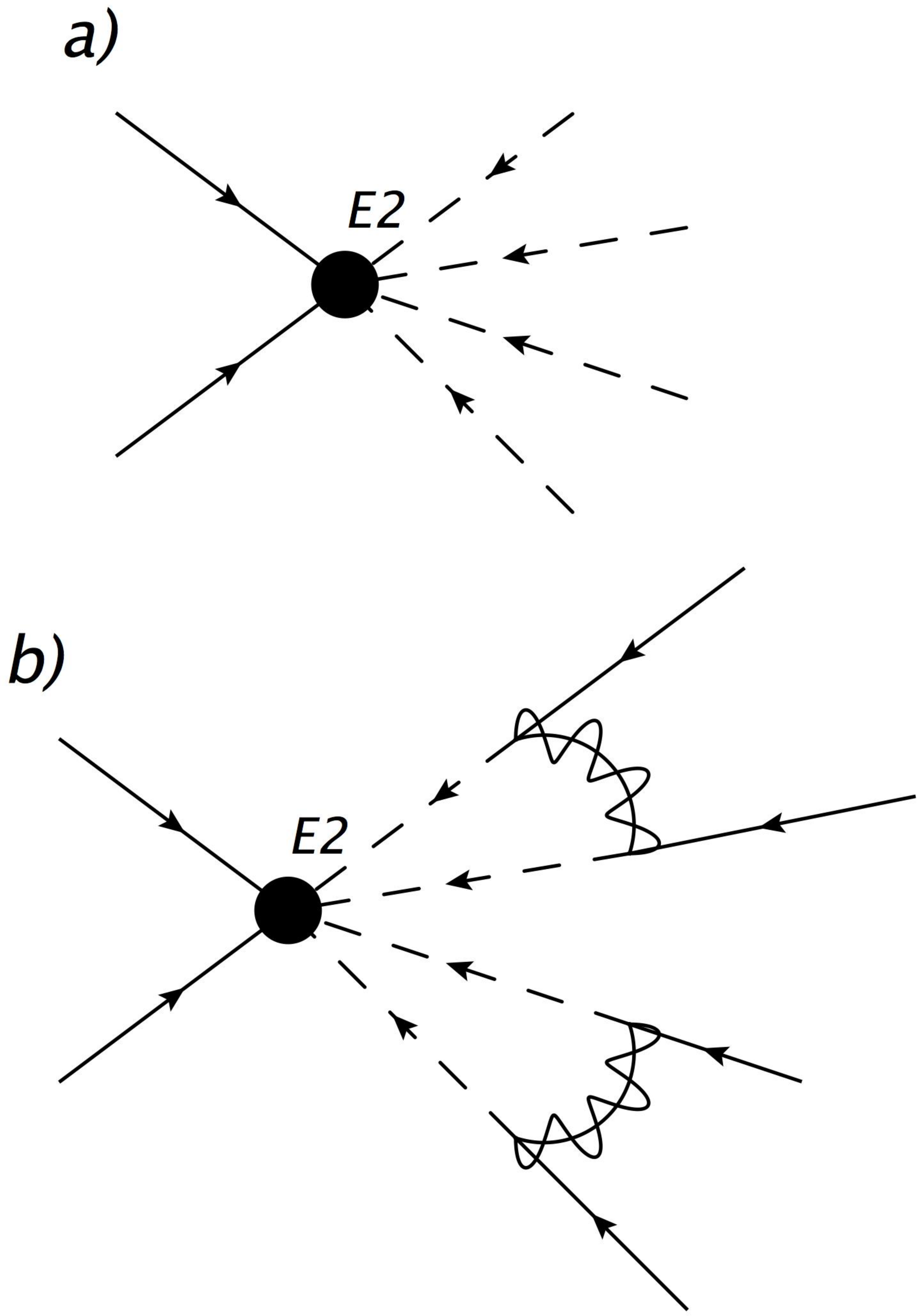}}
\vspace*{-1ex}
\caption{a)
Four anti-squark production in high energy collisions of two quarks
$q_{f_{1}}q_{f_{2}}\rightarrow \bar{\tilde{q}}_{f_{3}}\bar{\tilde{q}}_{f_{4}}\bar{\tilde{q}}_{f_{5}}\bar{\tilde{q}}_{f_{6}}$.
b) Four quark production in in high energy collisions of two quarks
$q_{f_{1}}q_{f_{2}}\rightarrow \bar{q}_{f_{3}}\bar{q}_{f_{4}}\bar{q}_{f_{5}}\bar{q}_{f_{6}}$.
Black vertices are induced by exotic instantons.  }
\label{plot}   % \ref{plot}
\end{figure}

 An operator $\mathcal{O}_{\Lambda\bar{\Lambda}}$ generates not only $\Lambda-\bar{\Lambda}$ transition
 but also $NN\rightarrow KK$ transition. As discussed in the previous section
 $\delta m_{\Lambda \bar{\Lambda}}$ can be constrained up to
 $10^{-6}\, \rm eV$ by next generation of experiments.
 However, $NN\rightarrow KK$ is just constraining $\mathcal{O}_{\Lambda\bar{\Lambda}}$
 up to $\delta m_{\Lambda \bar{\Lambda}}<10^{-21}\, \rm eV$.
 $\delta m_{\Lambda \bar{\Lambda}}$ is related to a New Physics scale
by
 \be{NPs}
 \delta m_{\Lambda \bar{\Lambda}}\simeq y_{2}\frac{\Lambda_{QCD}^{6}}{\mathcal{M}_{\Lambda\bar{\Lambda}}^{5}}\,,
 \ee
 so that $\mathcal{M}_{\Lambda\bar{\Lambda}}\simeq 100\, \rm TeV$ scale.
 On the other hand, $\mathcal{O}_{n\bar{n}}$ is actually constrained
by $\delta m_{n\bar{n}}<10^{-23}\, \rm eV$, corresponding to
 $\mathcal{M}_{n\bar{n}}>300\, \rm TeV$ \cite{NNbar0a,NNbar0b}.
 The next generations of experiments will test \footnote{See \cite{Buchoff:2015qwa} for a discussion on perturbative renormalization corrections of $n-\bar{n}$ operators.} $\mathcal{M}_{n\bar{n}}\simeq 1000\, \rm TeV$ \cite{NNbar3}.
 In our model,
 $$\frac{\mathcal{M}_{\Lambda\bar{\Lambda}}}{\mathcal{M}_{n\bar{n}}}=\frac{y_{2}}{y_{1}}=10^{-18}\div 10^{18}\,.$$
Such a large hierarchy range is naturally  understood by
 couplings arising from
  mixed disk amplitudes.
 For example the number $10^{-18}$ is understood
 as $\frac{k^{(1)}_{1}k^{(1)}_{1}k^{(2)}_{1}k^{(2)}_{1}k^{(2)}_{2}k^{(2)}_{2}}{k^{(1)}_{1}k^{(1)}_{1}k^{(2)}_{1}k^{(2)}_{1}k^{(2)}_{1}k^{(2)}_{1}}\simeq 10^{-3}\times 10^{-3}\times 10^{-3} \times 10^{-3}\times 10^{-3}\times 10^{-3}$.
  As a consequence, a $100\, \rm TeV$-scale test of $\Lambda-\bar{\Lambda}$
 is motivated from the theoretical side, independently from $n\leftrightarrow\bar{n}$ limits.
 A $100\, \rm TeV$ scale scenario
 for $\Lambda-\bar{\Lambda}$ can be naturally obtained with
 $e^{+S_{E2}}\simeq 1$ (small 3-cycles wrapped by the $E2$-brane),
  $M_{S}\simeq M_{SUSY}\simeq 100\, \rm TeV$.
  In this case, a final test-bed for this model can be provided
  by future proton-proton $100\, \rm TeV$ colliders beyond LHC.
  In fact a $uds\rightarrow \bar{u}\bar{d}\bar{s}$ transition
  will be directly tested, mediated by an exotic instanton in collision.
 However, a more intriguing scenario can be opened in the case
 of $10\, \rm TeV<M_{SUSY}\simeq M_{S}<100\, \rm TeV$.
 In this case, planned proton-proton colliders can test other flavor amplitudes
 like
 $ud\rightarrow \bar{c}\bar{s}\bar{s}\bar{s}, \bar{c}\bar{s}\bar{b}\bar{b}$
that are less constrained by $n-\bar{n}$ or $NN\rightarrow \pi\pi,KK$ processes.
In particular, the predicted experimental processes for a high energy proton-proton collider
are $pp\rightarrow 4\tilde{q}$.
This leads to several different channels.
This is the case of $pp\rightarrow 4q+4\chi^{0}$
leading to four jets and missing transverse energy $pp\rightarrow 4j+E_{M.T}$
or $pp\rightarrow 3j+t+E_{M.T}$ (with standard top decays) and so on.
Other interesting channels are coming from
stops productions and successive decays like $\tilde{t}\rightarrow W+\tilde{b}\rightarrow W+b+\chi^{0}$,
$\tilde{t}\rightarrow b+\chi^{+}\rightarrow b+W+\chi^{0}$.

As discussed in subsection A, cross sections of these processes have a peculiar behavior not common to
gauge models because of their exponential decrease up to $\Lambda$ for $s>>\Lambda^{2}$.

\section{Conclusions}

In this paper, we have discussed phenomenological implications
of  a new class of instantons
known as exotic instantons.
They can generate $\Delta B=2$ violating transitions
as $n\leftrightarrow \bar{n}, \Lambda\leftrightarrow \bar{\Lambda}$ and $\Delta B=2$ high energy collisions
like $qq\rightarrow \bar{\tilde{q}}\bar{\tilde{q}}\bar{\tilde{q}}\bar{\tilde{q}}, \bar{q}\bar{q}\bar{q}\bar{q}$
--in hypercharge preserving combinations.
We have explored the possibility to detect exotic instantons in future colliders,
in comparison with present low energy limit channels like
$n\leftrightarrow \bar{n}, NN\rightarrow \pi\pi,KK$.

We summarize our main conclusions as follow:

i) contrary to other non-perturbative solutions like electroweak gauge instantons,
exotic instantons can induce effective operator with a high coupling.
As a consequence, their effects can be seen in low energy observables
as well as in high energy colliders.

ii) A neutron-antineutron transition can be generated by exotic instantons
and it can provide an indirect test-bed for a class of models
mentioned in this paper
\footnote{This is an example of a different UV completion of six quarks
operator from a non-perturbative classical configurations, rather
than extra heavy fields. This has intriguing analogies with
classicalization
\cite{Dvali1,Dvali2,Dvali3,Dvali4,Addazi:2015ppa,CBH1}.
Reference \cite{Addazi:2015ppa} is related to discussions in \cite{Addazi:2015dxa}.
}.

iii) Our model predicts  $\Lambda-\bar{\Lambda}$ transitions.
The possibility to test these after
in future high luminosity electron-positron colliders seems
very far from our present technological possibility,
if compared to the actual related limits from $NN\rightarrow KK$ transitions.

iv) $\Delta B=2$ exotic instantons can be reached in the next generation of
high energy proton-proton colliders beyond LHC, well compatible with neutron-antineutron limits,
$NN\rightarrow \pi\pi,KK$ and so on.
We stressed how cross-section running with Center of Mass energy cannot
be reproduced by any quantum field theory model.
In fact an exponential softening of the cross-section cannot be
reproduced from any other UV completion of the
six quark effective operator in context quantum field theories.
This is a feature distinguishing our string theory model
from other quantum field theory ones.

v) Cosmological impact of exotic instantons revealed in \cite{Addazi:2015goa,Addazi:2015yna} provides
additional constraints on their parameters. Differently from the
case of electroweak gauge instantons, due to the enhancement of the
effect of exotic instantons in high energy collisions, a direct
quantitative relationship between cosmological and physical consequences
of the model considered is possible.

The class of models suggested here strongly motivates
two directions for future experimental physics:
neutron-antineutron experiments
and
high-energy colliders beyond LHC.
Eventually, a detection of exotic instanton-mediated processes can motivate the construction
of technologically challenging high luminosity collider in order to detect
a $\Lambda-\bar{\Lambda}$ transition or new rare physics experiments
searching for $NN\rightarrow KK$.
 These measurements could constrain the
exotic instantons' geometry and their intersections (with ordinary D-branes) and their
wrapped 3-cycles on the $CY_{3}$.
Future beyond LHC (such as the proposed CEPC+SppC) might render us new exciting surprises in
higher-energy and higher-luminosity frontiers.

\vspace{1cm}

{\large \bf Acknowledgments}
\vspace{3mm}

We would like to thank M.~ Bianchi, Z.~Berezhiani, G.~Dvali,
A.~Dolgov, C~Gomez and O.~Kanchelli, for useful discussions on these
aspects. The author XWK also thanks Hai-Bo Li for helpful advice.
A.~A work was supported in part by the MIUR research grant
"Theoretical Astroparticle Physics" PRIN 2012CPPYP7. XWK's work is
partly supported by the DFG and the NSFC through funds provided to
the Sino-German CRC 110 ¡°Symmetries and the Emergence of Structure
in QCD¡±when he was in J\"ulich, and by the Ministry of Science and
Technology of R.O.C. under Grant No. 104-2112-M-001-022 when he
works in Taiwan from April. The work by MK was performed within the
framework of the Center FRPP supported by MEPhI Academic Excellence
Project (contract 02.03.21.0005, 27.08.2013), supported by the
Ministry of Education and Science of Russian Federation, project
3.472.2014/K and grant RFBR 14-22-03048.

\section*{Appendix A: superpotential calculus.}

In our paper,
we have considered
an instanton with a rigid $O(1)$ symmetry in context of IIA superstring theory.
This instanton has the
universal zero mode structure $dx^{4}d^{2}\theta$
which yields a new term to the
holomorphic F-terms in the effective supergravity
action.
The computation of these effects can be
done from the Conformal Field Theory prospective.
In particular, one can compute the supepotential
of an M-point correlator in a string instantonic background
$\langle \Phi_{a_{1}b_{1}} ...\Phi_{a_{M}b_{M}} \rangle_{\mathcal{E}}$
--where $\mathcal{B}$ denoted the instantonic background and
$\Phi$ are generic physical fields in the bi-fundamental representations
of two generic $N'$ and $M'$ stacks of D6-branes

The correlator is related to the effective
supergravity quantities in the action as follows
\cite{Cvetic:2007ku,Blumenhagen:2009qh}:
 \be{PHIPHI}
\langle \Phi_{a_{1}b_{1}}...\Phi_{a_{N}b_{N}}\rangle_{\mathcal{B}}=\frac{e^{\mathcal{K}/2}Y_{\Phi_{a_{1}b_{1}},...,\Phi_{a_{M}b_{M}}}}{\sqrt{K_{a_{1}b_{1},...,a_{M}b_{M}}}}\,,
 \ee
 where $\mathcal{K}$ is the non-holomorphic Kh\"aler potential,
 $K$ is the Kh\"aler metric and $Y$ is the holomorphic superpotential coupling.
The following generic formula for the computation of the correlation function
in semiclassical approximation is
\cite{Cvetic:2007ku,Blumenhagen:2009qh}:
 \be{PHIPHI2}
\langle \Phi_{a_{1}b_{1}},...,\Phi_{a_{M}b_{M}}\rangle_{\mathcal{B}}\simeq \int d^{4}xd^{2}\theta \sum_{conf}
\prod_{a}(\prod_{i=1}^{I^{+}}d\lambda_{a}^{i})(\prod_{i=1}^{I^{-}}d\bar{\lambda}_{a}^{i})
 \ee
 $$\exp(-S^{(0)})\exp\left( \sum_{b}C_{\mathcal{E}\mathcal{D}_{b}^{*}}+C_{\mathcal{E}O^{*}}\right)$$
$$ \times
 \langle \hat{\Phi}_{a_{1}b_{1}}[\vect{x}_{1}]\rangle_{\lambda_{a_{1}}\bar{\lambda}_{b_{1}}}... \langle \hat{\Phi}_{a_{L}b_{L}}[\vect{x}_{L}]\rangle_{\lambda_{a_{L}}\bar{\lambda}_{b_{L}}}\,,$$
 where $\hat{\Phi}_{a_{k}b_{k}}$ is the chain-product of all the the vertex operators
 at fixed value of $k$
 $$\hat{\Phi}_{a_{k}b_{k}}=\Phi_{a_{k}x_{k,1}}\cdot \Phi_{x_{k,1}x_{k,2}}\cdot ...\cdot \Phi_{x_{k,n-1},x_{k,n}}\cdot \Phi_{x_{k,n},b_{k}}\,$$
while
$$\langle \hat{\Phi}_{a_{1}b_{1}}[\vect{x}_{1}]\rangle_{\lambda_{a_{1}},\bar{\lambda}_{b_{1}}}$$
a CFT disk correlator for $\hat{\Phi}$
and for the charged zero modes $\lambda_{a_{1}},\bar{\lambda}_{b_{1}}$
inserted in the boundary of the mixed disk amplitudes.
$C_{\mathcal{E}\mathcal{D}_{b}^{*}},C_{\mathcal{E}O^{*}}$
denote the sum all over the annulus
diagrams with and without one cross-cap.

From Eqs.(\ref{PHIPHI})-(\ref{PHIPHI2}),
the holomorphic superpotential coupling can be entirely
re-expressed in terms of holomorphic couplings in
the CFT amplitudes:
\be{CFTHO}
Y_{\Phi_{a_{1}b_{1}}...\Phi_{a_{M}b_{m}}}=\sum_{conf.}\exp(-S_{0})
\ee
$$\times Y_{\lambda_{a_{1}}\hat{\Phi}_{a_{1}b_{1}}[\vect{x}_{1}],\bar{\lambda}_{b_{1}}}
...Y_{\lambda_{a_{1}}\hat{\Phi}_{a_{L}b_{L}}[\vect{x}_{L}]\bar{\lambda}_{b_{L}}}\,,$$
where $S_{0}$ corresponds to the vacuum disk amplitude for the E2-brane:
$$S_{\mathcal{E}}^{(0)}=-\langle 1 \rangle_{Disc}=\frac{1}{g_{s}}\frac{V_{E2}}{l_{s}^{3}},$$
with $V_{E2}$ is volume of the 3-cycles wrapped by the E2-instanton.
This formula generalizes the prescription
given in Section II,
in the particular case of brane intersections described above.
In particular, the instanton can reproduce the
six quark operator if and only if with the number of branes intersections considered above.

\end{document}